\documentclass{INTERSPEECH2023}
\usepackage{subcaption}
\usepackage{pifont}
\newcommand{\xmark}{\ding{55}}
\newcommand{\cmark}{\ding{51}}

\interspeechcameraready

\title{Careful Whisper - leveraging advances in automatic speech recognition for robust and interpretable aphasia subtype classification}
\name{Laurin Wagner$^1$, Mario Zusag$^1$, Theresa Bloder$^1$}
\address{
  $^1$nyra health GmbH, Austria}
\email{lwagner@nyra.health, mzusag@nyra.health, tbloder@nyra.health}

\begin{document}

\maketitle

\begin{abstract}
This paper presents a fully automated approach for identifying speech anomalies from voice recordings to aid in the assessment of speech impairments. By combining Connectionist Temporal Classification (CTC) and encoder-decoder-based automatic speech recognition models, we generate rich acoustic and clean transcripts. We then apply several natural language processing methods to extract features from these transcripts to produce prototypes of healthy speech. Basic distance measures from these prototypes serve as input features for standard machine learning classifiers, yielding human-level accuracy for the distinction between recordings of people with aphasia and a healthy control group. Furthermore, the most frequently occurring aphasia types can be distinguished with $90\%$ accuracy. The pipeline is directly applicable to other diseases and languages, showing promise for robustly extracting diagnostic speech biomarkers.
\end{abstract}

\section{Introduction}
Aphasia is defined as an acquired language disorder following brain injury \cite{ivanova2022aphasia}, for instance, after stroke or traumatic head injury. Following a stroke, the most frequently occurring aphasia types are global, anomic, Wernicke’s, and Broca’s aphasia \cite{pedersen2004aphasia}. Kirshner and Wilson \cite{kirshner2021aphasia} summarized the characteristics according to these syndromes. In Broca's aphasia, the speech pattern is nonfluent, often referred to as “agrammatic” or “telegraphic”. Patients with Broca’s aphasia often make phonemic errors that are inconsistent from utterance to utterance. In contrast, the speech pattern of patients with Wernicke’s aphasia is effortless, sometimes even overly fluent, containing verbal paraphasias, neologisms, and jargon productions. Auditory comprehension is impaired, which often renders patients unconscious to their nonsense speech productions. Global aphasia might be considered a summation of the symptoms observed in Broca’s and Wernicke’s aphasia, characterized by nonfluent speech paired with poor comprehension skills. Finally, in anomic aphasia, the main deficit concerns word access or retrieval from the mental lexicon which is characterized by pauses and circumlocutions. The distinction between these four aphasia types is, however, not always that clear cut and, several mixed aphasia types exist. In clinical practice, therapeutic strategies to approach any language-related deficits have to be aligned with the particular symptoms to maximize rehabilitative potential \cite{landrigan2021data}. The accurate classification of aphasia is therefore highly relevant. However, the process of correctly evaluating these subtleties of speech and language can be quite time-consuming. Employing automated approaches on speech recordings with interpretable insights would enable clinicians to initiate speech therapy more quickly, thus expediting rehabilitation.

Several attempts have been made to classify different aphasia types and other speech pathologies from manual transcripts \cite{fraser2014automated, fromm2022enhancing}. Fraser et al. \cite{fraser2014automated} conducted a study with the objective of discerning between individuals diagnosed with semantic dementia, progressive nonfluent aphasia and healthy controls, utilizing features extracted from manual transcripts for syntactic complexity and measures of fluency. Fromm et al. \cite{fromm2022enhancing} employed the CHAT format manual transcriptions \cite{macwhinney2000childes} to extract features on subtasks of AphasiaBank \cite{macwhinney2011aphasiabank} data and study k-means clusters of these features. Other works have developed approaches to detect aphasia automatically from spontaneous speech \cite{le2018automatic} \cite{aphasia2019automatic}. Notably, Chatzoudis et al. \cite{zeroshotAphasia} use the XLSR-53 model from Conneau et al. \cite{xlsr}, which was pre-trained on multiple languages using the contrastive self-supervised task and wav2vec2.0 architecture from Baevski et al. \cite{wav2vec2} to extract language-invariant linguistic features for classifying aphasia. These models learn high-quality audio representations using Transformer encoder blocks \cite{transformer} and are successively fine-tuned for downstream tasks. To fine-tune these models, Conneau et al. \cite{xlsr} have added a classifier representing the output vocabulary of the respective downstream task on top of the XLSR-53 model by training on labeled datasets using a Connectionist Temporal Classification (CTC) loss \cite{graves2006connectionist}. 

Even though the underlying wav2vec2.0 architecture contains multiple Transformer layers, which can encode contextual information, fine-tuned XLSR-53 models without an additional language model tend to produce more literal, acoustic outputs than the encoder-decoder Whisper model from Radford et al. \cite{whisper}, therefore leaving filler words and recurring utterances as artifacts. Whisper was trained in a supervised fashion on weakly-supervised large-scale raw text labels without significant text standardization, such that the decoder part of Whisper works as an audio-conditioned language model and eliminates many filler words, recurring utterances and other artifacts. We propose to combine the output of these two models and apply carefully selected feature extraction methods on aligned and annotated transcripts to build prototypes of healthy speech. These prototypes are used to normalize features for inference in order to create interpretable insights for anomalous speech recordings. 

The contributions of our work are as follows:
\begin{itemize}
    \item With a fully automated approach, we achieve human-level accuracy on distinguishing speech recordings of aphasic patients from a healthy control group. 
    \item To the best of our knowledge, it is the first approach exploiting the different decoding mechanisms of CTC and encoder-decoder based methods to create rich annotations for automatic feature extraction.
    \item With a fully automated approach, we achieve over 90\% accuracy on classifying speech between a healthy control group, anomic, Wernicke’s and Broca’s aphasia. 
    \item We provide several unseen scores for improving the accuracy of automatically assessing speech anomalies.
\end{itemize}

\section{Proposed method}

In this section, we describe the proposed pipeline for creating robust speech characteristics for assessing speech impairments. The pipeline consists of a prototype construction and classification phase. Both phases use feature extraction methods in three successive steps with the difference that the prototype phase uses extracted features from a population of healthy speech samples to estimate healthy speech distributions, while the classification phase uses distance measures from these distributions to generate classifiable features. We have used the 89 hours of free speech samples from the Europarl speech translation corpus \cite{europarl-dataset} to construct the healthy prototypes.

\begin{figure*}[t]
  \centering
  \includegraphics[width=\textwidth]{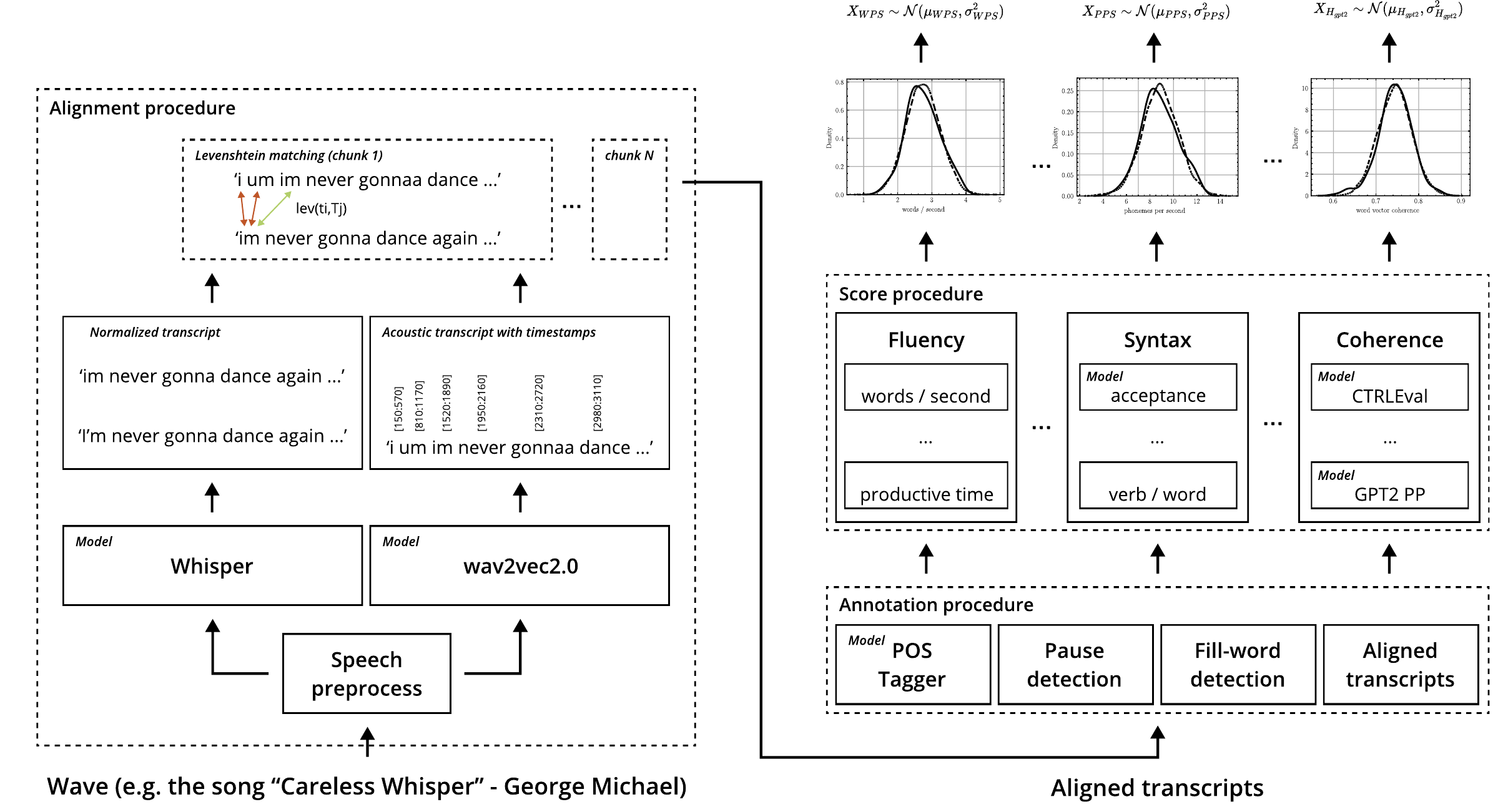}
  \caption{Alignment, annotation and score procedure}
  \label{fig:prototype_procedure}
\end{figure*}

The three successive steps consist of first constructing acoustic and clean transcripts leveraging the speech recognition models XLSR-53 and Whisper on speech chunks as described in Section \ref{section:alignment-pipeline}. We have used the large version of XLSR-53 with 24 Transformer encoder blocks and the large version of Whisper. Second, we enrich the aligned transcripts with fully-automated annotations for pauses, filler words, and part-of-speech (POS) tags. Third, we are computing scores for coherence, fluency, syntax, lexical richness, and pronunciation as described in Section \ref{section:scores-pipeline}. For classification, speech recordings are fed into the alignment, annotation and score pipeline to compute the distance between each score and the score prototypes. These distances are then used for training simple machine learning classifiers and are used for the experiments in Section \ref{section:experiments}. The steps for computing prototypes are portrayed in Figure \ref{fig:prototype_procedure}.

\subsection{Alignment and annotation}
\label{section:alignment-pipeline}

The whole pipeline operates directly on speech waves, which are sampled at \SI{16}{\kilo\hertz}. The speech wave is fed into Whisper to produce syntactically correct transcripts containing punctuations, removing single utterances, repetitions and other artifacts, and into XLSR-53 to produce the more literal and standardized outputs. In the following, we will call the decoded XLSR-53 output \textit{acoustic} and the Whisper output \textit{clean} transcripts. Due to the configuration of wav2vec2.0, tokens are decoded with a \SI{20}{\milli\second} stride resulting in precise timing information on token level and by extension on word level \cite{ctc_segmentation}. Whisper provides timings on a sentence level only. 

We combine the strengths of these two transcripts by aligning their standardized output via the minimization of the Levenshtein distance \cite{levenshtein1966binary} as portrayed in Figure \ref{fig:prototype_procedure}. This enables us to identify unmatched words or utterances and tokens as potential filler words or recurring utterances and allows us to augment the clean transcripts with accurate timings for the matched tokens from the acoustic transcripts in a fully-automated fashion. 

The recordings were segmented into contiguous speech chunks, each containing a minimum of 200 words, based on clean transcriptions. To ensure accurate syntax scores, we confirmed that the final word in each chunk corresponded to the end of a sentence, with Whisper providing punctuation marks. The aligned acoustic, accurate timings, and clean text data were used in our annotation process, illustrated in Figure \ref{fig:prototype_procedure}. In this step, we are augmenting the aligned transcripts with POS tags using RoBERTa from Liu et al. \cite{roberta}, detect pauses from silence tokens with more than \SI{300}{\milli\second} of duration and filler words from timings and unmatched utterances. These augmented transcripts are then fed into our score pipeline, as described in Section \ref{section:scores-pipeline}.

\subsection{Scores for language abilities}
\label{section:scores-pipeline}
The goal of the proposed scores is to capture multiple aspects of language abilities as shown in Table \ref{tab:scores}. We accumulate the distribution of each score  and approximate it with a normal distribution, since they fit the empiric distributions well and serve as prototypes for healthy speech.

\begin{table}[h!]
  \caption{Scores for identifying speech characteristics.}
  \label{tab:scores}
  \centering
  \begin{tabular}{ r@{}l  l }
    \toprule
    \multicolumn{2}{l}{\textbf{Language ability}} & 
                                         \multicolumn{1}{l}{\textbf{Feature}} \\
    \midrule
    \multicolumn{2}{l}{Fluency} & \multicolumn{1}{l}{Words and phonemes per second} \\
    &  & Percentage time spoken ~~~   \\
    &  & Productive time ratio ~~~  \\
    &  & Pause length and distance ~~~             \\
    &  & Pause per word ~~~           \\
    &  & Mean phoneme length nouns  ~~~    \\

   \midrule
   \multicolumn{2}{l}{Lexical richness} & \multicolumn{1}{l}{TTR, MATTR} \\
    &  & gzip ratio ~~~          \\
    &  & HD-D, MTLD  ~~~              \\
    &  & Word information ~~~    \\
    \midrule
   \multicolumn{2}{l}{Syntax} & 
                                         \multicolumn{1}{l}{Noun, verb and adjective ratio} \\
    &  & Grammar acceptance  ~~~             \\
    &  & Mean sentence length  ~~~             \\

    \midrule
   \multicolumn{2}{l}{Pronunciation} & 
                                         \multicolumn{1}{l}{Word error rate acoustic} \\
    &  & Character error rate acoustic ~~~             \\
    \midrule
    \multicolumn{2}{l}{Coherence} & 
                                         \multicolumn{1}{l}{CTRLEval} \\
    &  & Word vector coherence ~~~             \\
    &  & GPT2 perplexity ~~~             \\

    \bottomrule
  \end{tabular}
  
\end{table}

\subsubsection{Fluency}

The \textbf{words per second} score is computed on clean transcripts, while \textbf{percentage time spoken} uses the aggregated time of the acoustic transcripts divided by the total time of the provided audio chunk. For measuring the \textbf{productive time ratio}, we divide the strict spoken time (discounting pauses) of the acoustic transcript by the strict spoken time of the clean transcript. We count every successive set of silence tokens from acoustic transcripts that are longer than \SI{300}{\milli\second} as a pause to compute \textbf{pause length} scores. Sampling all pause lengths as a distribution gives insights into the frequency and duration of speech breaks. We use different quantiles (10, 25, 50, 75, 95) of this distribution as features. \textbf{Pause distance} by contrast computes the time between successive pauses and again, samples the pause distances as a distribution and uses quantiles (10, 25, 50, 75, 95) as features. The \textbf{pause per word} score is defined as the ratio of pauses longer than \SI{300}{\milli\second} and the total number of words from the clean transcript. We also included two phoneme scores, for which we use the automatic phonemizer from Bernard and Titieux \cite{phonemizer} to transform the clean and acoustic text into phonemes. For \textbf{mean phoneme length nouns}, we aggregate and average the lengths of all words tagged as nouns from clean transcripts. Finally, the \textbf{phonemes per second} score is computed on the phonemized version of clean transcripts.  

\subsubsection{Lexical richness}

All lexical richness scores are computed on clean transcripts. We have computed \textbf{type token ratio} (TTR) and \textbf{MATTR} \cite{MATTR} with window sizes of 10, 25 and 50 words. Using the intuition that redundant text is easier to compress, we have used the gzip algorithm, creating \textbf{gzip ratio} as a measurement for redundancy, dividing the length of compressed by uncompressed byte string representations. Further we use \textbf{HD-D} and \textbf{MTLD} as described in \cite{HD-D}, which were proven to capture lexical diversity well, even under varying text length. \textbf{Word information} is defined as the entropy of the normalized token count distribution of the lemmatized tokens \cite{WIM}.

\subsubsection{Syntax}

All syntax scores are computed on clean transcripts. \textbf{Noun, verb and adjective ratio} are simply the ratio of the specified POS tags to the number of words in the clean transcript. The \textbf{grammar acceptance} feature is defined as the softmax probability of a RoBERTa-large model \cite{roberta} fine-tuned for grammar acceptability on the CoLA Corpus \cite{CoLA-Corpus}. Finally, we use the \textbf{mean sentence length} as an indicator of noticeable short or sprawling phrases. 

\subsubsection{Pronunciation}
The \textbf{word/character error rate acoustic} scores are the word/character error rates between the acoustic and clean transcripts, intuitively capturing speech intelligibility proportional to the diverging outputs of Whisper and XLSR-53.

\subsubsection{Coherence}
All coherence scores are evaluated on Whisper transcripts. \textbf{CTRLEval} is a reference-free metric to assess the coherence for controlled text generation tasks as detailed by Ke et al. \cite{ctrleval}. For \textbf{word vector coherence}, we use the cosine similarity to assess the local coherence between averaged word vectors of neighboring sentences. Word vectors are taken from spaCy's \cite{spaCy} en\_core\_web\_sm corpus. At last, with the \textbf{GPT2 perplexity} score, we evaluate the perplexity from GPT2 by Radford et al. \cite{radford2019language} with a sliding window size of 512 tokens using the standard GPT2 tokenizer.

\section{Experiments and results}
\label{section:experiments}

\subsection{Experimental setup}
\label{section:experiment_setup}

To validate our approach, we are using speech samples extracted from the AphasiaBank database \cite{macwhinney2011aphasiabank} for English. AphasiaBank comprises a collection of interviews between clinicians and subjects afflicted with aphasia, as well as healthy control subjects. These interviews are transcribed in the CHAT transcription format  \cite{macwhinney2000childes} and include timestamps for interviewer and patient speech segments. Most interviews contain a label indicating which form of aphasia the subject is suffering from. There are $272$ individuals from the control group, $136$ with anomic, $80$ with Broca's, $60$ conduction, $29$ with Wernicke's, $9$ with transmotor, $3$ individuals with global aphasia and $35$ individuals who do not exhibit symptoms of aphasia as determined by the Western Aphasia Battery (WAB-R) \cite{WAB-R}. For $348$ subjects the (WAB-R) Aphasia Quotient (AQ) is included. We remove all speech uttered by the interviewer as given by the timestamps from the CHAT protocols and concatenate the resulting audio clips, which are input to our pipeline. For each speech chunk, we compute the scores from Section \ref{section:scores-pipeline} as outlined in Figure \ref{fig:prototype_procedure} and average multiple score vectors from the same interview. The distance for each averaged score $s_i$ to the healthy prototype of this score, estimated by $ X \sim \mathcal{N}(\mu_s,\,\sigma_s^{2})$, are then used as features for classification and computed by $s_i = \frac{\sigma_{s}}{|\mu_{s} - s_i|}$, if $|s_i - \mu_{s}| > \sigma_{s}$ else it is set to 1.

To ensure comparability with the results from Chatzoudis et al. \cite{zeroshotAphasia}, we adopt a Leave-One-Subject-Out (LOSO) cross-validation strategy for all classification experiments. We use a standard linear Support Vector Classifier SVC \cite{cortes1995support} with regularization parameter $C=0.1$ and max\_iter $=50000$ for classification. We do not use task-specific feature selection and train on all of our features. All experiments were run on a virtual machine using an NVIDIA-A100 GPU.

\subsection{Results}
\label{section:results}
\subsubsection{Aphasia versus Control}

To facilitate a direct comparison with Chatzoudis et al. \cite{zeroshotAphasia}, we have included the classification results of our approach for aphasic (including all subtypes) versus healthy control patients in Table \ref{aphasia_vs_control}. The outcome of this analysis reveals that our fully automatic method exhibits superior performance relative to the one of \cite{zeroshotAphasia}, even when comparing ourselves to features extracted directly from the manually annotated CHAT transcriptions. Furthermore, since we classify on averaged chunks of similar length, instead of the full transcript, we are not benefiting from the discriminatory information provided by total word output. These findings provide compelling evidence of the robustness and efficacy of our fully-automatic approach for real-world applications. 

\begin{table}[th]
  \caption{Classification metrics for aphasia versus healthy control group using linguistic features on AphasiaBank.}
  \label{aphasia_vs_control}
  \centering
  \begin{tabular}{lcll}
    \toprule
    \multicolumn{1}{l}{\textbf{Method}} & 
    \multicolumn{1}{c}{\textbf{Manual annotation}} & 
    \multicolumn{1}{l}{\textbf{Accuracy}} &
    \multicolumn{1}{l}{$\mathbf{F_1}$} \\ 
    \midrule
    SVC \cite{zeroshotAphasia} & \xmark  & $93.9$ & - \\
    SVC \cite{zeroshotAphasia} & \cmark & $97.4$  & - \\
    SVC \cite{optimal_transport_aphasia_detection} & \cmark & - & $85.9$ \\
    \midrule
    \textbf{Ours (SVC)} & \xmark & $\mathbf{98.6}$ & $\mathbf{98.5}$    \\
    \bottomrule
  \end{tabular}
\end{table}

\subsubsection{Aphasia subtype classification}
\label{subsection:aphasia_subtype}
To validate the effectiveness and robustness of the feature extraction pipeline, we use a linear SVC to differentiate three types of aphasia and the healthy control group as a multi-class classification problem. We have used three of the four main aphasia subtypes, i.e. anomic, Wernicke’s, and Broca’s aphasia. We exclude global aphasia since there are only 3 people with this diagnosis in AphasiaBank. Further details on the classification performance are summarized in Table \ref{aphasia_subtype_classification_metrics}.

\begin{table}[th!]
  \caption{Classification metrics for a 4-class classification problem differentiating between aphasia subtypes and the control group of the AphasiaBank dataset.}
  \label{aphasia_subtype_classification_metrics}
  \centering
  \begin{tabular}{llll}
    \toprule
    \multicolumn{1}{l}{\textbf{Class}} & 
    \multicolumn{1}{l}{\textbf{Precision}} &
    \multicolumn{1}{l}{\textbf{Recall}} & 
    \multicolumn{1}{l}{$\mathbf{F_1}$}\\
    \midrule
    Control  & $97.5$ & $98.5$ & $98.0$ \\
    Anomic  & $83.1$ & $86.8$ & $84.9$ \\ 
    Broca  & $80.0$ & $75.0$ & $77.5$ \\ 
    Wernicke  & $91.7$ & $68.8$ & $78.6$ \\ 
    \midrule
    Weighted average  & $90.6$ & $90.6$ & $90.6$ \\
    
    \bottomrule
  \end{tabular}
\end{table}

\subsubsection{Classification results per score category}
Each score category should contribute meaningful information to the multi-class subtype classification problem from section \ref{subsection:aphasia_subtype}. To this end, we trained a linear SVC that only utilizes the subset of scores from the respective score category to differentiate the three aphasia subtypes and the healthy control group. The results for the weighted average precision, recall and $F_1$ scores for the aphasia subtypes and control group per score category are condensed in Table \ref{aphasia_subtype_classification_per_category}. 

\begin{table}[th]
  \caption{Weighted average precision, recall and $F_1$ scores for a 4-class classification problem differentiating between aphasia subtypes and the control group of the AphasiaBank dataset using only features from a single score category.}
  \label{aphasia_subtype_classification_per_category}
  \centering
  \begin{tabular}{llll}
    \toprule
    \multicolumn{1}{l}{\textbf{Score Category}} & 
    \multicolumn{1}{l}{\textbf{Precision}} & 
    \multicolumn{1}{l}{\textbf{Recall}} & 
    \multicolumn{1}{l}{$\mathbf{F_1}$}\\
    \midrule
    Fluency  & $81.8$ & $82.0$ & $81.6$ \\
    Coherence  & $68.9$ & $71.0$ & $69.1$ \\
    Lexical Richness  & $76.5$ & $77.9$ & $76.5$ \\
    Syntax  & $76.9$ & $78.3$ & $77.4$ \\
    Pronunciation  & $61.7$ &  $67.3$ & $63.9$ \\
    \bottomrule
  \end{tabular}
\end{table}

\subsubsection{WAB-R AQ Regression}
The automatic estimation of the aphasia quotient (AQ) from spontaneous speech has numerous potential benefits, including progress monitoring without the need for frequent repetition of the WAB-R assessment procedures. We follow the evaluation procedure from Le et al. \cite{WAB_regression} and frame the WAB-R AQ prediction as a regression problem, with our proposed score distances as features and the AQ scores as the target labels for the $348$ individuals, for which the AQ was included. The results are portrayed in Table \ref{wab-r-aq-pred}.

\begin{table}[th]
  \caption{WAB-R AQ Prediction. MAE=Mean Absolute Error, PC = Pearsons's Correlation, SVR = Support Vector Regression}
  \label{wab-r-aq-pred}
  \centering
  \begin{tabular}{lcll}
    \toprule
    \multicolumn{1}{l}{\textbf{Method}} & 
    \multicolumn{1}{c}{\textbf{Manual annotation}} & 
    \multicolumn{1}{l}{\textbf{PC}} &
    \multicolumn{1}{l}{\textbf{MAE}} \\ 
    \midrule
    SVR \cite{WAB_regression} & \xmark  & $0.776$ & $9.90$ \\
    SVR \cite{WAB_regression} & \cmark  & $0.787$ & $9.54$ \\
    \midrule
    \textbf{Ours (SVR)} & \xmark & $\mathbf{0.815}$ & $\mathbf{8.19}$    \\
  \end{tabular}
\end{table}

\section{Conclusions}
The proposed approach showed high level of robustness for identifying aphasia and aphasic subtypes from free speech recordings and presents a pipeline that can be adapted to other diseases and languages with minimal effort. We used the same pipeline to evaluate the proposed set of features for a dementia classification challenge, beating the baseline by a large margin. In the near future, we want to investigate the distance measures from healthy prototypes to build accurate and interpretable speech biomarkers for multiple diseases. Also, the optimal number of words per chunk for practical applications, the transferability of features across multiple languages and the correlation between multiple features need further investigation. 
\newpage
\bibliographystyle{IEEEtran}
\bibliography{mybib}

\begin{thebibliography}{10}
\providecommand{\url}[1]{#1}
\csname url@samestyle\endcsname
\providecommand{\newblock}{\relax}
\providecommand{\bibinfo}[2]{#2}
\providecommand{\BIBentrySTDinterwordspacing}{\spaceskip=0pt\relax}
\providecommand{\BIBentryALTinterwordstretchfactor}{4}
\providecommand{\BIBentryALTinterwordspacing}{\spaceskip=\fontdimen2\font plus
\BIBentryALTinterwordstretchfactor\fontdimen3\font minus
  \fontdimen4\font\relax}
\providecommand{\BIBforeignlanguage}[2]{{%
\expandafter\ifx\csname l@#1\endcsname\relax
\typeout{** WARNING: IEEEtran.bst: No hyphenation pattern has been}%
\typeout{** loaded for the language `#1'. Using the pattern for}%
\typeout{** the default language instead.}%
\else
\language=\csname l@#1\endcsname
\fi
#2}}
\providecommand{\BIBdecl}{\relax}
\BIBdecl

\bibitem{ivanova2022aphasia}
M.~V. Ivanova and N.~F. Dronkers, ``Aphasia: How our language system can
  “break”,'' \emph{Frontiers for young minds}, vol.~10, 2022.

\bibitem{pedersen2004aphasia}
P.~M. Pedersen, K.~Vinter, and T.~S. Olsen, ``Aphasia after stroke: type,
  severity and prognosis,'' \emph{Cerebrovascular diseases}, vol.~17, no.~1,
  pp. 35--43, 2004.

\bibitem{kirshner2021aphasia}
H.~S. Kirshner and S.~M. Wilson, ``Aphasia and aphasic syndromes,''
  \emph{Bradley's Neurology in Clinical Practice E-Book}, vol. 133, 2021.

\bibitem{landrigan2021data}
J.-F. Landrigan, F.~Zhang, and D.~Mirman, ``A data-driven approach to
  post-stroke aphasia classification and lesion-based prediction,''
  \emph{Brain}, vol. 144, no.~5, pp. 1372--1383, 2021.

\bibitem{fraser2014automated}
K.~C. Fraser, J.~A. Meltzer, N.~L. Graham, C.~Leonard, G.~Hirst, S.~E. Black,
  and E.~Rochon, ``Automated classification of primary progressive aphasia
  subtypes from narrative speech transcripts,'' \emph{cortex}, vol.~55, pp.
  43--60, 2014.

\bibitem{fromm2022enhancing}
D.~Fromm, J.~Greenhouse, M.~Pudil, Y.~Shi, and B.~MacWhinney, ``Enhancing the
  classification of aphasia: a statistical analysis using connected speech,''
  \emph{Aphasiology}, vol.~36, no.~12, pp. 1492--1519, 2022.

\bibitem{macwhinney2000childes}
B.~MacWhinney, ``The childes project: Tools for analyzing talk: Volume i:
  Transcription format and programs, volume ii: The database,'' 2000.

\bibitem{macwhinney2011aphasiabank}
B.~MacWhinney, D.~Fromm, M.~Forbes, and A.~Holland, ``Aphasiabank: Methods for
  studying discourse,'' \emph{Aphasiology}, vol.~25, no.~11, pp. 1286--1307,
  2011.

\bibitem{le2018automatic}
D.~Le, K.~Licata, and E.~M. Provost, ``Automatic quantitative analysis of
  spontaneous aphasic speech,'' \emph{Speech Communication}, vol. 100, pp.
  1--12, 2018.

\bibitem{aphasia2019automatic}
Y.~Qin, T.~Lee, and A.~P.~H. Kong, ``Automatic assessment of speech impairment
  in cantonese-speaking people with aphasia,'' \emph{IEEE journal of selected
  topics in signal processing}, vol.~14, no.~2, pp. 331--345, 2019.

\bibitem{zeroshotAphasia}
\BIBentryALTinterwordspacing
G.~Chatzoudis, M.~Plitsis, S.~Stamouli, A.-L. Dimou, A.~Katsamanis, and
  V.~Katsouros, ``Zero-shot cross-lingual aphasia detection using automatic
  speech recognition,'' 2022. [Online]. Available:
  \url{https://arxiv.org/abs/2204.00448}
\BIBentrySTDinterwordspacing

\bibitem{xlsr}
A.~Conneau, A.~Baevski, R.~Collobert, A.~Mohamed, and M.~Auli, ``Unsupervised
  cross-lingual representation learning for speech recognition,'' \emph{arXiv
  preprint arXiv:2006.13979}, 2020.

\bibitem{wav2vec2}
\BIBentryALTinterwordspacing
A.~Baevski, H.~Zhou, A.~Mohamed, and M.~Auli, ``wav2vec 2.0: A framework for
  self-supervised learning of speech representations,'' 2020. [Online].
  Available: \url{https://arxiv.org/abs/2006.11477}
\BIBentrySTDinterwordspacing

\bibitem{transformer}
A.~Vaswani, N.~Shazeer, N.~Parmar, J.~Uszkoreit, L.~Jones, A.~N. Gomez,
  {\L}.~Kaiser, and I.~Polosukhin, ``Attention is all you need,''
  \emph{Advances in neural information processing systems}, vol.~30, 2017.

\bibitem{graves2006connectionist}
A.~Graves, S.~Fern{\'a}ndez, F.~Gomez, and J.~Schmidhuber, ``Connectionist
  temporal classification: labelling unsegmented sequence data with recurrent
  neural networks,'' in \emph{Proceedings of the 23rd international conference
  on Machine learning}, 2006, pp. 369--376.

\bibitem{whisper}
\BIBentryALTinterwordspacing
A.~Radford, J.~W. Kim, T.~Xu, G.~Brockman, C.~McLeavey, and I.~Sutskever,
  ``Robust speech recognition via large-scale weak supervision,'' 2022.
  [Online]. Available: \url{https://arxiv.org/abs/2212.04356}
\BIBentrySTDinterwordspacing

\bibitem{europarl-dataset}
\BIBentryALTinterwordspacing
J.~Iranzo-Sánchez, J.~A. Silvestre-Cerdà, J.~Jorge, N.~Roselló, A.~Giménez,
  A.~Sanchis, J.~Civera, and A.~Juan, ``Europarl-st: A multilingual corpus for
  speech translation of parliamentary debates,'' 2019. [Online]. Available:
  \url{https://arxiv.org/abs/1911.03167}
\BIBentrySTDinterwordspacing

\bibitem{ctc_segmentation}
\BIBentryALTinterwordspacing
L.~Kürzinger, D.~Winkelbauer, L.~Li, T.~Watzel, and G.~Rigoll,
  ``{CTC}-segmentation of large corpora for german end-to-end speech
  recognition,'' in \emph{Speech and Computer}.\hskip 1em plus 0.5em minus
  0.4em\relax Springer International Publishing, 2020, pp. 267--278. [Online].
  Available: \url{https://doi.org/10.1007%2F978-3-030-60276-5_27}
\BIBentrySTDinterwordspacing

\bibitem{levenshtein1966binary}
V.~I. Levenshtein \emph{et~al.}, ``Binary codes capable of correcting
  deletions, insertions, and reversals,'' in \emph{Soviet physics doklady},
  vol.~10, no.~8.\hskip 1em plus 0.5em minus 0.4em\relax Soviet Union, 1966,
  pp. 707--710.

\bibitem{roberta}
\BIBentryALTinterwordspacing
Y.~Liu, M.~Ott, N.~Goyal, J.~Du, M.~Joshi, D.~Chen, O.~Levy, M.~Lewis,
  L.~Zettlemoyer, and V.~Stoyanov, ``Roberta: A robustly optimized bert
  pretraining approach,'' 2019. [Online]. Available:
  \url{https://arxiv.org/abs/1907.11692}
\BIBentrySTDinterwordspacing

\bibitem{phonemizer}
\BIBentryALTinterwordspacing
M.~Bernard and H.~Titeux, ``Phonemizer: Text to phones transcription for
  multiple languages in python,'' \emph{Journal of Open Source Software},
  vol.~6, no.~68, p. 3958, 2021. [Online]. Available:
  \url{https://doi.org/10.21105/joss.03958}
\BIBentrySTDinterwordspacing

\bibitem{MATTR}
M.~A. Covington and J.~D. McFall, ``Cutting the gordian knot: The
  moving-average type--token ratio (mattr),'' \emph{Journal of quantitative
  linguistics}, vol.~17, no.~2, pp. 94--100, 2010.

\bibitem{HD-D}
P.~M. McCarthy and S.~Jarvis, ``Mtld, vocd-d, and hd-d: A validation study of
  sophisticated approaches to lexical diversity assessment,'' \emph{Behavior
  research methods}, vol.~42, no.~2, pp. 381--392, 2010.

\bibitem{WIM}
K.~T. Cunningham and K.~L. Haley, ``Measuring lexical diversity for discourse
  analysis in aphasia: Moving-average type--token ratio and word information
  measure,'' \emph{Journal of Speech, Language, and Hearing Research}, vol.~63,
  no.~3, pp. 710--721, 2020.

\bibitem{CoLA-Corpus}
A.~Warstadt, A.~Singh, and S.~R. Bowman, ``Neural network acceptability
  judgments,'' \emph{Transactions of the Association for Computational
  Linguistics}, vol.~7, pp. 625--641, 2019.

\bibitem{ctrleval}
\BIBentryALTinterwordspacing
P.~Ke, H.~Zhou, Y.~Lin, P.~Li, J.~Zhou, X.~Zhu, and M.~Huang, ``{CTRLE}val: An
  unsupervised reference-free metric for evaluating controlled text
  generation,'' in \emph{Proceedings of the 60th Annual Meeting of the
  Association for Computational Linguistics (Volume 1: Long Papers)}.\hskip 1em
  plus 0.5em minus 0.4em\relax Dublin, Ireland: Association for Computational
  Linguistics, May 2022, pp. 2306--2319. [Online]. Available:
  \url{https://aclanthology.org/2022.acl-long.164}
\BIBentrySTDinterwordspacing

\bibitem{spaCy}
M.~Honnibal, I.~Montani, S.~Van~Landeghem, and A.~Boyd, ``{spaCy:
  Industrial-strength Natural Language Processing in Python},'' 2020.

\bibitem{radford2019language}
A.~Radford, J.~Wu, R.~Child, D.~Luan, D.~Amodei, I.~Sutskever \emph{et~al.},
  ``Language models are unsupervised multitask learners,'' \emph{OpenAI blog},
  vol.~1, no.~8, p.~9, 2019.

\bibitem{WAB-R}
A.~Kertesz, ``Western aphasia battery--revised,'' 2007.

\bibitem{cortes1995support}
C.~Cortes and V.~Vapnik, ``Support-vector networks,'' \emph{Machine learning},
  vol.~20, pp. 273--297, 1995.

\bibitem{optimal_transport_aphasia_detection}
\BIBentryALTinterwordspacing
A.~Balagopalan, J.~Novikova, M.~B.~A. McDermott, B.~Nestor, T.~Naumann, and
  M.~Ghassemi, ``Cross-language aphasia detection using optimal transport
  domain adaptation,'' 2019. [Online]. Available:
  \url{https://arxiv.org/abs/1912.04370}
\BIBentrySTDinterwordspacing

\bibitem{WAB_regression}
D.~Le, K.~Licata, and E.~M. Provost, ``Automatic quantitative analysis of
  spontaneous aphasic speech,'' \emph{Speech Communication}, vol. 100, pp.
  1--12, 2018.

\end{thebibliography}

\end{document}